\documentclass[prb,aps,twocolumn,showpacs,amsmath,amssymb,floatfix,citeautoscript]{revtex4}
\usepackage{graphicx}
\usepackage{dcolumn}
\usepackage{bm}
\usepackage{amsmath}
\usepackage{amssymb}
\usepackage[normal]{subfigure}

\makeatletter

\usepackage{color}

\begin{document}

\title{Thermalization mechanism for time-periodic finite isolated interacting quantum systems}

\author{Dong E. Liu}
\affiliation{Department of Physics and Astronomy, Michigan State University, East Lansing, MI 48824, USA
}

\date{\today}

\begin{abstract}
We present a theory to describe thermalization mechanism for time-periodic
finite isolated interacting quantum systems.  The long time 
asymptote of natural observables in Floquet states is  directly related to 
averages of these observables governed by a time-independent effective Hamiltonian. 
We prove that if the effective system is nonintegrable and satisfies eigenstate 
thermalization hypothesis,  quantum states of such time-periodic isolated systems 
will thermalize. After a long time evolution, system will relax to a stationary state, 
which only depends on an initial energy of the effective Hamiltonian and follows a 
generalized eigenstate thermalization hypothesis. A numerical test for the periodically 
modulated Bose-Hubbard model, with the extra nearest neighbor interaction on the 
bosonic lattice, agrees with the theoretical predictions.
\end{abstract}


\maketitle

\section{Introduction and motivation}

Time-periodic quantum mechanics can be successfully understood 
in the framework of the Floquet theorem \cite{shirley65,sambe73}. Putting this approach into the rigorous footing to describe thermodynamics of time-periodic correlated quantum systems has long been an elusive goal due to great complications. Analytical and numerical studies of open Floquet systems based on relatively simple models revealed highly nontrivial behaviors, and also left a great number of open problems \cite{Kohler97,Hone97,Breuer00,Kohn01,Hone09,ketzmerick10,arimondo12}. Recent technical developments in cold-atomic physics \cite{Kinoshita06,Hofferberth07}, photonic lattices \cite{Peleg07,Bahat-Treidel08},
and graphene \cite{Novoselov05,GrapheneRMP} provide experimental platforms to investigate existing proposals and resurge theoretical interests. Perhaps the most appealing potential application of Floquet systems is in alternative ways of realizing and controlling exotic quantum states~\cite{inoue10,kitagawa10,lindner11,kitagawaGF11,jiang11,Dahlhaus11,reynoso12,
Liu13,RudnerPRX13,Kundu13,RechtsmanNature13,WangScience13}.  However, the problem of ordering of the Floquet spectrum is still unresolved, which is very important when considering effects of interactions in a closed system or coupling to the external bath. Therefore, understanding thermodynamics of the Floquet systems becomes a key to characterize emerging topological phases. 

Time-evolution of isolated integrable Floquet system shows that observables in a continuous Floquet spectrum limit tend to a time-periodic steady state \cite{Russomanno12}. The long time evolution can be described by a Floquet version of Generalized Gibbs Ensemble \cite{Lazarides13}. For interacting systems~\cite{DAlessio&Rigol14}, the quasi-energy level statistics shows properties described by a circular ensemble of random matrices for different driving frequencies, except for a crossover regime. In the thermodynamic limit,
periodic driven ergodic systems after long-time evolution will relax to distribution with infinite temperature \cite{DAlessio&Rigol14,Lazarides14,Ponte14}, and periodic driven many-body localized (non-ergodic) systems have memory of the initial state \cite{Ponte14}. Similar infinite temperature distributions were also found numerically in certain open driven systems \cite{Breuer00,Hone09,ketzmerick10}, which is connected to chaotic regimes of those systems \cite{ketzmerick10}.

Here, we focus on a finite isolated Floquet quantum systems. We argue that the physical observables in such systems can be investigated in a framework of an effective Hamiltonian. Based on such an approach, we show that the long time evolution of a natural observable in Floquet system can be obtained by the time evolution of a time-independent system. For finite interacting system, we assume that the quasi-energy spectrum has high density, but the degeneracy of the quasi-energies are excluded or only occurs occasionally. Then, if the effective Hamiltonian is nonintegrable and satisfies eigenstate thermalization hypothesis (ETH) \cite{Deutsch91,Srednicki94,Rigol08}, it can be shown that an isolated Floquet quantum system will  thermalize. After a long time evolution, the system will reach a stationary state characterized by a (microcanonical) finite temperature ensemble of the effective Hamiltonian. The stationary state only depends on an initial energy, this initial energy and the ordering of the Floquet states should only be understood in the effective Hamiltonian system. For an example, we consider finite driven interacting bosonic lattice with periodic modulation, and study the relaxation of momentum distribution function. We numerically demonstrate that the momentum distribution will relax to their microcanonical predictions based on the generalized ETH.

\section{Observable expectation values for Floquet systems}

Let us consider a time-periodic Hamiltonian $H(t)=H(t+T)$. Based on Floquet theorem \cite{shirley65,sambe73}, the corresponding Schrodinger equation has a complete set of time-periodic solutions $|\Phi_{\alpha}(t)\rangle=e^{-i\epsilon_{\alpha}t}|\phi_{\alpha}(t)\rangle$ with $|\psi_{\alpha}(t)\rangle=|\phi_{\alpha}(t+T)\rangle $, that satisfy $[H(t)-i\partial_t]|\phi_{\alpha}(t)\rangle=\epsilon_{\alpha}|\phi_{\alpha}(t)\rangle$ (hereafter $\hbar=1$). Here $\epsilon_{\alpha}$ and $|\phi_{\alpha}(t)\rangle$ are called quasi-energies and Floquet states. The Floquet states, whose quasi-energies differ only by an integer multiples of $\omega=2\pi/T$, describe the same physical states, and thus, one can choose $\epsilon_{\alpha}\in [0,\omega]$.

Assume we can find a time dependent unitary transformation $U_F(t)=U_F(t+T)$ with $U_F(t=0)=\mathbb{I}$ ($\mathbb{I}$ is identity) such that the effective Hamiltonian after the transformation becomes time independent
\begin{equation}
H_{\rm eff}=U_F(t) H(t) U_F^{\dagger}(t)+i\partial_t\Big( U_F(t)\Big)U_F^{\dagger}(t).
\end{equation}
The Floquet states and quasi-energies are related to the time independent problem as
$|\phi_{\alpha}(t)\rangle=U_F^{\dagger}(t)|\phi_{\alpha 0}\rangle$, $\epsilon_{\alpha}=\mod(\widetilde{\epsilon}_{\alpha},\omega)$
and $H_{\rm eff} |\phi_{\alpha 0}\rangle=\widetilde{\epsilon}_{\alpha}|\phi_{\alpha 0}\rangle$.
Therefore, the matrix elements for an observable $A$ in Floquet basis can be written as
\begin{equation}
\langle \phi_{\alpha}(t)| A | \phi_{\beta}(t)\rangle = \langle \phi_{\alpha 0}| A_{\rm eff}(t) | \phi_{\beta 0}\rangle,
\end{equation}
where $A_{\rm eff}(t) = U_F(t) A U_F^{\dagger}(t) = \sum_{n} e^{-i n \omega t} A_{\rm eff}^{[n]}$. 
If the period $T\ll\tau_{\mathrm{in}}$ is small compared to the inelastic scattering time $\tau_{\mathrm{in}}$ the experimentally relevant quantities are their time average over a full period
\begin{equation}
\frac{1}{T}\int_0^T dt \langle \phi_{\alpha}(t)| A | \phi_{\beta}(t)\rangle = \langle \phi_{\alpha 0}| A_{\rm eff}^{[0]} | \phi_{\beta 0}\rangle.
\end{equation}
The physical observables in time-periodic system can be thus investigated in the frame of an effective Hamiltonian. The exact $U_F(t)$ can only be found for some special models. For general cases, an approximated transformation can be found by a rotating frame transformation \cite{eckardt05} in the large driving frequency limit $\omega\gg D$ (where $D$ is the spectral width of an effective Hamiltonian [e.g. $D=\mathrm{max}[J,U,U_N]$ in Eq.(\ref{eq:Hamiltonian})]. Beyond this limit, but still under the condition $\omega>D$, the effective Hamiltonian and effective observables can be obtained by a flow equation method \cite{KehreinBook,Verdeny13} up to a certain order in $D/\omega$.

\section{Thermalization of a time periodic isolated quantum systems}

In generic isolated nonintegrable quantum systems without periodic modulation, many macroscopic quantities will tend to stationary values after long time evolution, i.e. thermalize \cite{PolkovnikovRMP11}. This thermalization behavior can be described by a so-called eigenstate thermalization hypothesis (ETH) \cite{Deutsch91,Srednicki94,Rigol08}. The expectation value of a nature observable $\langle \Psi_{\alpha}|A|\Psi_{\alpha}\rangle$ in an energy $E_{\alpha}$ 
eigenstate $|\Psi_{\alpha}\rangle$ of a large nonintegrable many-body system equals the microcanonical average at the average energy $E_{\alpha}$: $\langle \Psi_{\alpha}|A|\Psi_{\alpha}\rangle=\langle A \rangle_{\rm micro}(E_{\alpha})$. Therefore, the long time average of $\langle \Psi(t)|A|\Psi(t)\rangle$ with the initial condition $|\Psi(t=0)\rangle=\sum_{\alpha} C_{\alpha} |\Psi_{\alpha}\rangle$ is then \cite{PolkovnikovRMP11}
\begin{eqnarray}
\overline{\langle \Psi(t)|A|\Psi(t)\rangle} &=& \sum_{\alpha} |C_{\alpha}|^2 A_{\alpha\alpha}=\langle A \rangle_{\rm micro}(E_{0}) \nonumber\\
      &=&  \frac{1}{\mathcal{N}_{E_0,\Delta E}} \sum_{|E_0-E_{\alpha}|<\Delta E} A_{\alpha\alpha},
\label{eq:Heff_ETH}
\end{eqnarray}
where $A_{\alpha\beta}=\langle \Psi_{\alpha}|A|\Psi_{\beta}\rangle$, $E_0=\sum_{\alpha} |C_{\alpha}|^2 E_{\alpha}$ is the energy of the initial state, and $\mathcal{N}_{E_0,\Delta E}$ is the number of eigenstates with energies in the window $[E_0-\Delta E, E_0+\Delta E]$ and $\overline{\cdots}=\mathrm{lim}_{\tau\rightarrow \infty}(1/\tau)\int_0^{\tau}\cdots dt$.

For finite systems with periodic modulation, the natural questions are: whether Floquet states thermalize,
what is the mechanism of Floquet state thermalization, and whether the stationary
state depends only on a few parameters? The time evolution of a many-body state for time-periodic $H(t)$ can be written as a linear combination of Floquet states:c$|\Psi(t)\rangle=\sum_{\alpha} C_{\alpha} e^{-i\epsilon_{\alpha}t}|\phi_{\alpha}(t)\rangle$. 
We will assume 1) the quasi-energy spectrum has high density, 
2) the quasi-energies are non-degenerate, 
or the degeneracy only occurs occasionally. Note that the second
condition is not true in thermodynamic limit, where quasi-energies are all highly degenerate. For the specified conditions, the long time average of an observable $A$ is then
\begin{eqnarray}
\overline{\langle \Psi(t)|A|\Psi(t)\rangle}&=&\overline{\sum_{\alpha \beta} C_{\alpha}^{*} C_{\beta} e^{i(\epsilon_{\alpha}-\epsilon_{\beta})t}
                            \sum_{n} e^{-in\omega t} A_{\alpha\beta}^{[n]}}\nonumber\\                          
                           &\approx& \sum_{\alpha} |C_{\alpha}|^2 \frac{1}{T}\int_0^T dt \langle \phi_{\alpha}(t)| A | \phi_{\alpha}(t)\rangle\nonumber\\
                           &=& \sum_{\alpha} |C_{\alpha}|^2 \langle \phi_{\alpha 0}| A_{\rm eff}^{[0]} | \phi_{\alpha 0}\rangle.
\label{eq:Floquet_dynamics}
\end{eqnarray}
Therefore, the long time evolution of an observable in Floquet system will tend to a stationary value, and this value can be described by the effective Hamiltonian $H_{\rm eff}$ with eigenstates $\{| \phi_{\alpha 0}\rangle\}$ and eigenenergies $\{ \widetilde{\epsilon}_{\alpha} \}$ [note that quasi-energies are $\epsilon_{\alpha}=\mod(\widetilde{\epsilon}_{\alpha},\omega)$]. If we choose the condition $U_F(t=0)=\mathbb{I}$, one can make sure the initial state for both Floquet system and the effective system are the same: $|\phi_{\alpha}(t=0)\rangle=U_F(t=0)|\phi_{\alpha,0}\rangle=|\phi_{\alpha,0}\rangle$ and $|\Psi(t=0)\rangle=U_F(t=0)|\Psi_0(t=0)\rangle=|\Psi_0(t=0)\rangle$ (the subscript index $0$ is for the state in $H_{\rm eff}$),
and thus $\langle \Psi(t=0)|\phi_{\alpha}(t=0)\rangle=\langle \Psi_0(t=0)|\phi_{\alpha 0}\rangle=C_{\alpha}$.
Then, by combining the ETH shown in Eq.~(\ref{eq:Heff_ETH}) and the long time dynamics shown in Eq.~(\ref{eq:Floquet_dynamics}), we immediately obtain the main result of the paper: an observables of a time-periodic isolated nonintegrable quantum system will tend to a stationary state which only depends on the initial energy $E_0=\sum_{\alpha} |C_{\alpha}|^2 \widetilde{\epsilon}_{\alpha}$ from the effective Hamiltonian,
and follow a generalized ETH
\begin{eqnarray}
\overline{\langle \Psi(t)|A|\Psi(t)\rangle} &=& \sum_{\alpha} |C_{\alpha}|^2 \langle \phi_{\alpha 0}| A_{\rm eff}^{[0]} | \phi_{\alpha 0}\rangle \label{eq:Floquet_ETH}\\
      &=&  \frac{1}{\mathcal{N}_{E_0,\Delta E}} \sum_{|E_0-\widetilde{\epsilon}_{\alpha}|<\Delta E} \langle \phi_{\alpha 0}| A_{\rm eff}^{[0]} | \phi_{\alpha 0}\rangle.\nonumber
\end{eqnarray}
As the system size (or the number of particles) becomes increasingly larger, the degeneracy of quasi-energies occurs more frequently, and the stationary state after long-time evolution starts to deviate from the predictions above. In the thermodynamic limit, an infinite-temperature state is predicted in Refs.[\onlinecite{DAlessio&Rigol14,Lazarides14,Ponte14}].

\section{Driven bosonic interacting lattice} 

To study the relaxation and test the generalized ETH for the time-periodic isolated quantum system, we consider a modulated one-dimensional (1D) Bose-Hubbard model with an extra nearest neighbor interaction term
\begin{eqnarray}
H(t)&=&\sum_{j=1}^M \Big( J (c_j^{\dagger}c_{j+1}+h.c.)+\frac{U}{2} n_j (n_j -1) \nonumber\\
    && +U_N n_j n_{j+1}  \Big) +F(t) \sum_j j n_j
\label{eq:Hamiltonian}
\end{eqnarray}
where $c_j^{\dagger}$ ($c_j$) creates (annihilates) a boson at site$-j$ in the 1D chain, $n_j=c_j^{\dagger} c_j$, and the time-periodic
function $F(t)=F(t+T)$, for example, can be chosen as
\begin{equation}
F(t) = \left\{ 
  \begin{array}{l l}
    K & \quad \text{if $nT\leq t<(n+1/2)T$},\\
    -K & \quad \text{if $(n+1/2)T\leq t<(n+1)T$}.
  \end{array} \right.
\end{equation}
To obtain the effective Hamiltonian, we choose a time-dependent 
unitary transformation $U_F(t)=e^{-if(t)\sum_j j n_j}$ \cite{eckardt05}
with $df(t)/dt + F(t)=0$ and $f(t=0)=0$ so that
\begin{equation}
f(t) = \left\{ 
  \begin{array}{l l}
    -K(t-nT) & \; \text{if $nT\leq t<(n+\frac{1}{2})T$},\\
    K(t-(n+1)T) & \; \text{if $(n+\frac{1}{2})T\leq t<(n+1)T$}.
  \end{array} \right.
\end{equation}

\begin{figure}[t]
\centering
\includegraphics[width=3.2in,clip]{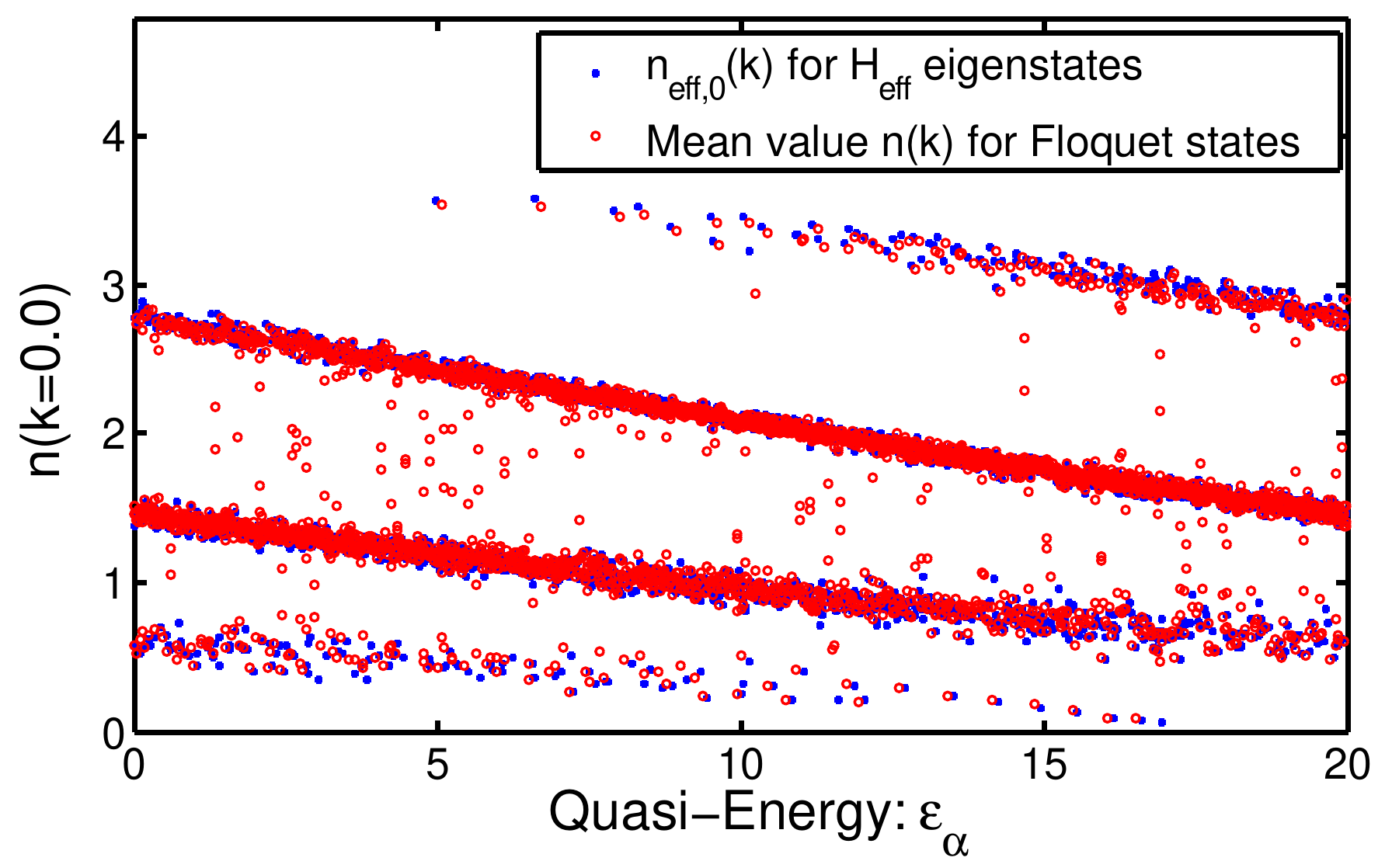}
\caption{(color online) The blue dots are the expectation values of
$n^{[0]}_{\mathrm{eff}}(k=0)$ for different eigenstates of $H_{\mathrm{eff}}$ plotted as a function of
$\rm mod (\widetilde{\epsilon_{\alpha}}, \omega)$, where $\widetilde{\epsilon_{\alpha}}$
are eigenenergies of  $H_{\mathrm{eff}}$. The red circles are the Floquet state mean value
of $n(k=0)$. We choose $J=2.4$, $U=0.8$, $U_N=0.7$, $\omega=20.0$, and $2\pi K/\omega=5.5$.
} 
\label{fig:NkCompare}
\end{figure}

By using this transformation and a further Fourier expansion, we can rewrite the Floquet Hamiltonian into an extended 
basis $| \{n_j\};m\rangle = e^{i f(t) \sum_j j n_j} e^{-i m \omega t}$,
\begin{eqnarray}
&& H^{F}_{m m'} = \langle\langle \{n_j\};m | H(t)-i\partial_t  |\{n_j\};m'\rangle\rangle \nonumber\\
        &&= \frac{1}{T} \int_0^T dt \bigg( e^{-i f(t) \sum_j j n_j} \Big[ \sum_{j=1}^M  J_0 (c_j^{\dagger}c_{j+1}+h.c.)\nonumber\\
        && +\frac{U}{2}\sum_j n_j (n_j -1)+U_N\sum_j n_j n_{j+1}+m'\omega \Big] \nonumber\\
        && e^{i f(t) \sum_j j n_j} \bigg) e^{i(m-m')\omega t},
\end{eqnarray}
where $\langle\langle\cdots\rangle\rangle=(1/T)\int_0^T dt \cdots $.
In an operator formalism (see also Ref. \onlinecite{Verdeny13}), the Hamiltonian becomes
\begin{eqnarray}
\hskip-.75cm
&&H^{F} = \sum_j \Big[ J (\mathcal{L}_{0}(K/\omega)c_j^{\dagger}c_{j+1}+h.c.)\nonumber\\
   \hskip-.75cm&& +\frac{U}{2} n_j (n_j -1)+U_N n_j n_{j+1} \Big] \otimes \mathbb{I} + \mathbb{I} \otimes \omega \hat{n}\nonumber\\
  \hskip-.75cm&& +\sum_{j,m\neq 0} 
             \Big[ \mathcal{L}_{m}(K/\omega) c_j^{\dagger}c_{j+1}+ \overline{\mathcal{L}}_{m}(K/\omega) c_{j+1}^{\dagger}c_{j} \Big] \otimes \hat{\sigma}_m
\end{eqnarray}
where $\mathcal{L}_{m} = (1/T)\int_0^T dt e^{im\omega t}e^{-if(t)}$ and  
$\overline{\mathcal{L}}_{m} = (1/T)\int_0^T dt e^{im\omega t}e^{if(t)}$ with $\mathcal{L}_{m}^{*}=\overline{\mathcal{L}}_{-m}$,
which are functions of only $K/\omega$.  The integer operator $\hat{n}$ is applied
to the vector $|n\rangle$ in Fourier spaces $\hat{n}|n\rangle=n|n\rangle$; and $\hat{\sigma}_m$ 
connects different Fourier components $\hat{\sigma}_m|n\rangle=|n+m\rangle$, and describe
the non-diagonal blocks in $H^F$. In this matrix form, diagonal blocks are separated 
by energy $\omega$, and the absolute value of prefactors $\mathcal{L}_{m}$ ($\overline{\mathcal{L}}_{m}$)
in non-diagonal blocks are less then unit. Therefore, in the large driving frequency limit $\omega\gg \{J, U, U_N\}$, the transitions between different diagonal blocks 
can be treated perturbatively.  Up to the leading order in $\mathrm{max}[J, U, U_N]/\omega$, one can neglect all
the non-diagonal blocks, and reach an effective Hamiltonian \cite{eckardt05}
\begin{equation}
H_{\mathrm{eff}}=\!\!\sum_j \!\Big[ J (\mathcal{L}_{0} c_j^{\dagger}c_{j+1}+h.c.) +\frac{U}{2} n_j (n_j -1)+U_N n_j n_{j+1} \Big].
\label{eq:EffectiveH}
\end{equation}
The higher order terms can be incorporated by using a flow equation method.
When the quasi-energy degeneracy occurs frequently, some corrections of $H_{\mathrm{eff}}$
can also be obtained by degenerate perturbation theory \cite{Eckardt&Holthaus08}.
Beyond those limits,  even though the theory of Eq.~(\ref{eq:Floquet_ETH}) is still valid, 
it is unclear how to obtain the effective Hamiltonian for interacting Floquet system. 
We note that in the limit $K\ll \omega$, one has $\mathcal{L}_{0}\rightarrow 1$ and thus
$H_{\mathrm{eff}}\approx (1/T)\int_0^T dt H(t)$ \cite{DAlessio&Rigol14}, in which modulation
plays no role at all.

\begin{figure}
\centering
\vspace{0.05in} 
\includegraphics[width=3.3in,clip]{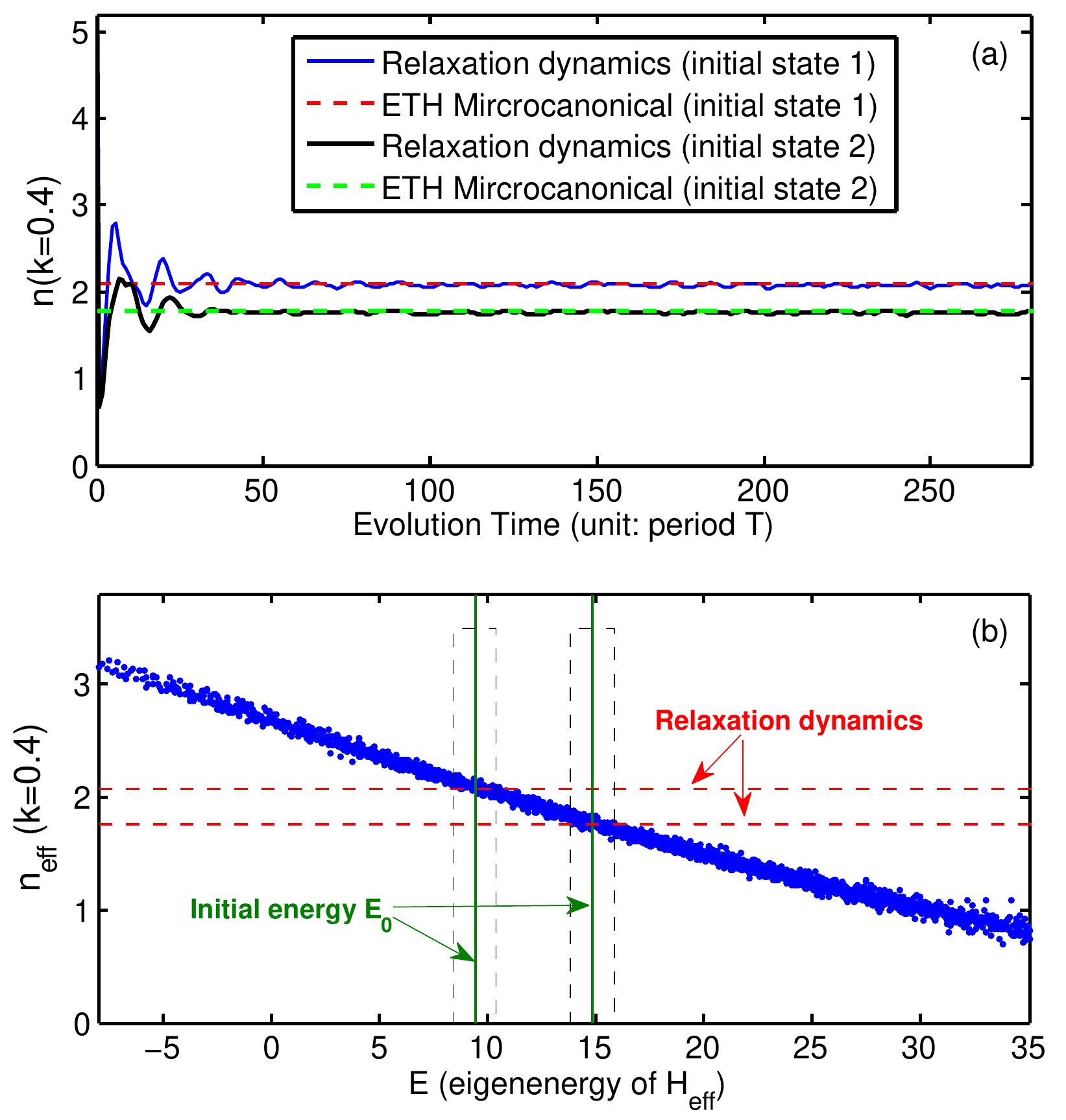}
\caption{(color online) (a) Solid line (blue for case-$1$, black for case-$2$): 
Relaxation dynamics of $n(k=0.4)$,
and each data point corresponds to the average the result over each full period.
Dashed-line (red for case-$1$, green for case-$2$): Result from generalized ETH for $n^{[0]}_{\mathrm{eff}}$ of $H_{\mathrm{eff}}$.
(b) Eigenstate expectation value $n^{[0]}_{\mathrm{eff}}$ as a function of eigenenergies
of $H_{\mathrm{eff}}$. The green solid line (left for case-$1$, right for case-$2$): the initial energy $E_0$. The red dashed line (upper for case-$1$, lower for case-$2$): long time relaxation dynamics. The black dashed box indicates the energy distribution of microcanonical ensemble in the calculation, 
i.e. $[E_0-\Delta E, E_0+\Delta E]$. Parameters are chosen 
the same as Fig. \ref{fig:NkCompare}.
} 
\label{fig:Relaxation}
\end{figure}

Now, we want to test the generalized ETH shown in Eq.~(\ref{eq:Floquet_ETH}). For our numerical
calculation, we choose the observable $A$ to be the momentum distribution function
\begin{equation}
n(k)=\frac{1}{M} \sum_{n=1}^M \sum_{m=1}^M e^{-2 i \frac{\pi k}{L}(r_n-r_m) } c_n^{\dagger} c_m 
\label{eq:nk_Ht}
\end{equation}
where the length $L=M a$ and $r_n = n a$ with lattice spacing $a$. 
The Fourier components of the effective momentum distribution are 
$n^{[p]}_{\mathrm{eff}}(k)=(1/T)\int_0^T dt e^{i p \omega t} U_F(t)n(k)U_F^{\dagger}(t)$.
Because of Eq.~(\ref{eq:Floquet_ETH}), we only need their zero component:
\begin{equation}
n^{[0]}_{\mathrm{eff}}(k)=\frac{1}{M} \sum_{n=1}^M \sum_{m=1}^M \mathcal{L}_0^{[n-m]} 
e^{-2 i \frac{\pi k}{L}(r_n-r_m) } c_n^{\dagger} c_m 
\label{eq:nk_Ht}
\end{equation}
where $\mathcal{L}_p^{[n-m]}=(1/T)\int_0^T dt e^{i p \omega t} e^{-i f(t)(n-m)}$.
Note that this is only correct up to the leading order for $H_{\mathrm{eff}}$.
If higher order terms, e.g. from flow equation method \cite{Verdeny13}, are included  in $H_{\mathrm{eff}}$,
one have to apply an extra rotation to $n(k)$.

\begin{figure}[b]
\centering 
\includegraphics[width=3.3in,clip]{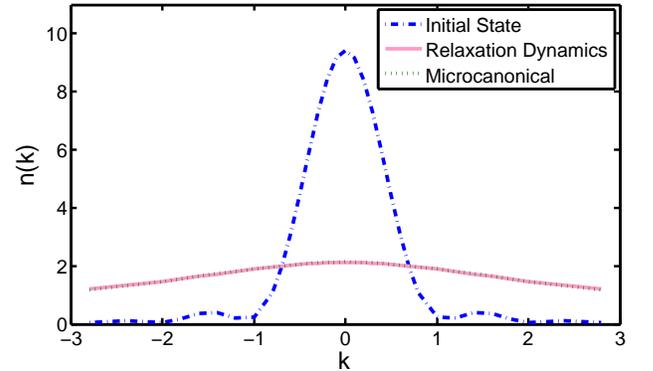}
\caption{(color online) The momentum distribution function in the inital
state case-$1$ (blue dot-dash line), after relaxation (pink solid line), and 
result from generalized ETH (green dotted line). Parameters are chosen 
the same as Fig. \ref{fig:NkCompare}.} 
\label{fig:Nkvskx}
\end{figure}

In the numerics, we consider $10$ bosons on a $6$-site chain, 
and choose the following 
parameters: $J=2.4$, $U=0.8$, $U_N=0.7$, $\omega=20.0$, and $2\pi K/\omega=5.5$.
We prepare the initial state $|\Psi(t=0)\rangle$ (two cases) to be the ground state of
the Bose-Hubbard model shown in Eq.~(\ref{eq:Hamiltonian}) (without modulation) but with the tilt potential
$-(1/41)\sum_j j n_j$ for the case-$1$, and $-(1/11)\sum_j j n_j$ for the case-$2$. 
In Fig. \ref{fig:NkCompare}, we compare the expectation values
of $n^{[0]}_{\mathrm{eff}}(k=0)$ for different $H_{\mathrm{eff}}$ eigenstates with the Floquet state mean value
of $n(k=0)$: $(1/T)\int_0^T dt \langle\phi_{\alpha}(t)|n(k=0)|\phi_{\alpha}(t) \rangle $.
The Floquet states $|\phi_{\alpha}(t) \rangle$ can be obtained by solving 
$U(t+T,t)|\phi_{\alpha}(t) \rangle = e^{-i \epsilon_{\alpha} T} |\phi_{\alpha}(t) \rangle$.
The agreement shown in the figure indicates that up to 
leading order the effective Hamiltonian  is sufficiently accurate for the current set of parameters. Those dissociated points between different 
bands come from the non-diagonal blocks in $H^{F}$ and the fact that
the eigenvectors are more sensitive to the parameter $K/\omega$. 

\begin{figure}[t]
\centering 
\includegraphics[width=3.2in,clip]{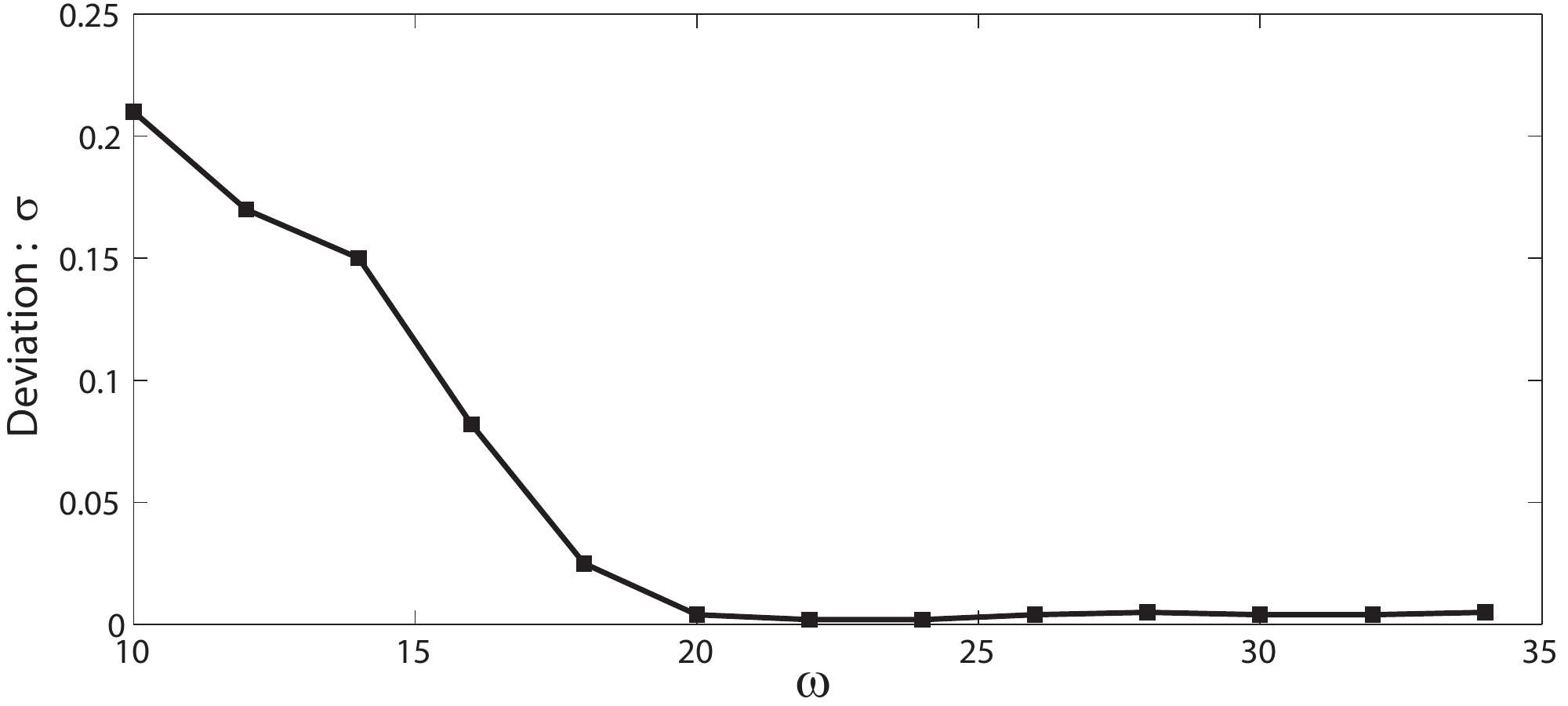}
\caption{The derviation $\sigma=|(n_{\rm R}-n_{\rm GETH})/n_{\rm R}|$ as a function of driving frequency $\omega$.
The parameters are the same as the case-1 in Fig. \ref{fig:Relaxation} (a): $k=0.4$,
$J=2.4$, $U=0.8$, $U_N=0.7$, $\omega=20.0$, and $2\pi K/\omega=5.5$. }
\label{fig:Deviation}
\end{figure}

Figure \ref{fig:Relaxation}(a) compares the relaxation dynamics of the momentum distribution
with the momentum distribution predicted by the generalized ETH. For relaxation dynamics,
we calculate the momentum distribution $n(k)$ in time evolution and average the result
over each full period. We consider $80$ steps in the time evolution for each period.
We can see that both initial states relax to their microcanonical predictions based on
a generalized ETH. Figure \ref{fig:Relaxation}(b) shows the eigenstate expectation value (EEV)
$\langle\phi_{\alpha 0}|n^{[0]}_{\mathrm{eff}}|\phi_{\alpha 0}\rangle$ as a function of eigenenergies
$\widetilde{\epsilon}_{\alpha}$ of $H_{\mathrm{eff}}$. The vertical green solid lines indicate
the value of initial energy $E_0$ for both initial states $1$ and $2$; 
the horizontal red dashed lines indicate the value of the relaxation dynamics.
The EEV does not fluctuate too much, and is almost a smooth
function of $\widetilde{\epsilon}_{\alpha}$ for most region. This fact explains why 
microcanonical predictions agree with the relaxation dynamics \cite{Deutsch91,Srednicki94,Rigol08}. Both crossing points are placed in the middle of the almost linear EEV data curves.
In Fig. \ref{fig:Nkvskx}, we plot the full momentum distribution $n(k)$ in the
initial state of case $1$, their relaxation dynamics, and 
the corresponding microcanonical predictions from the generalized ETH.
The plot indicates that the latter two agree with each other very well.
As the driving frequency $\omega$ becomes small, 
the effective Hamiltonian Eq. (\ref{eq:EffectiveH}) from high-frequency approximation is no longer valid,
and the non-diagonal blocks become important. In addition, for small driven frequency,
the quasi-energy spectrum become dense and the quasi-energy degeneracy may occur frequently. 
The deviation $\sigma=|(n_{\rm R}-n_{\rm GETH})/n_{\rm R}|$ due to small frequency effects are shown
in Fig. \ref{fig:Deviation}, where $n_{\rm R}$ is obtained from long time relaxation dynamics
and $n_{\rm GETH}$ is from generalized-ETH (microcanonical). Note that we fix the ratio between driving amplitude $K$
and the driving frequency , i.e. $2\pi K/\omega=5.5$, such that $H_{\mathrm{eff}}$ in Eq. (\ref{eq:EffectiveH})
is fixed, and thus does not approach the undriven Hamiltonian in large frequency limit.
For $\omega\gg J,U,U_N$ ,this deviation can be reduced
by including higher order terms in effective Hamiltonian. For general cases, computing the true 
effective Hamiltonian is a very hard task.

\section{Summary}

In summary, we propose and elaborate on the theoretical 
approach aiming to understand the thermalization and relaxation 
in finite Floquet isolated interacting quantum systems. 
If we assume 1) the quasi-energy spectrum has high density, 
2) the quasi-energies are non-degenerate, 
or the degeneracy only occurs occasionally.
The Floquet quantum states after long time evolution will
thermalize and follow a generalized ETH. 
The final stationary state only depends on the initial energy associated with the 
effective Hamiltonian; therefore, the ordering of the Floquet states (in
isolated interacting case) can only be understood by using the effective Hamiltonian.
Numerical tests in high frequency limit are carried out to compare the relaxation dynamics with the generalized ETH
for large modulation frequency limit, and show the agreement.
Beyond high frequency limit, calculating the effective Hamiltonian is nontrivial, and
the numerical test is a hard task.

\textit{Acknowledgment}. 
D.E.L. is grateful to A.Levchenko for valuable discussions and advice. This work was supported by NSF grant No. DMR-1401908.


\end{document}